\documentclass[twocolumn,superscriptaddress,aps,prb]{revtex4}

\usepackage{graphicx}

\begin{document}

\title{First-principles Design of Nanomachines}

\author{Jayanth R. Banavar}
\affiliation{Physics Department, Penn State University, 104 Davey Lab, University Park, PA 16802, USA}

\author{Marek Cieplak}
\affiliation{Institute of Physics, Polish Academy of Sciences,
Aleja Lotnikow 32/46, 02-668 Warsaw, Poland}

\author{Trinh Xuan Hoang}
\affiliation{Physics Department, Penn State University, 104 Davey Lab, University Park, PA 16802, USA}
\affiliation{Instititute of Physics, Vietnamese Academy of Science and Technology, 10 Dao Tan, Ba Dinh, Hanoi 10000, Vietnam}

\author{Amos Maritan}
\affiliation{Dipartimento di Fisica `G. Galilei', Universit\`a di Padova \& CNISM, unit\`a di Padova \& INFN, sezione di Padova,
Via Marzolo 8, 35131 Padova, Italy}


\begin{abstract}
Learning from nature's amazing molecular machines, globular proteins, we
present a framework for the predictive design of nano-machines.  We show
that the crucial ingredients for a chain molecule to behave as a machine
are its inherent anisotropy and the coupling between the local Frenet
coordinate reference frames of nearby monomers. We demonstrate that, even
in the absence of heterogeneity, protein-like behavior is obtained for a
simple chain molecule made up of just thirty hard spheres. This chain
spontaneously switches between two distinct geometries, a single helix and
a dual helix, merely due to thermal fluctuations.
\end{abstract}



\maketitle

Significant advances in laboratory techniques for tailoring and
processing materials at the atomic scale have resulted in
nanotechnology becoming an increasingly mature field with great
promise. One of the exciting goals of the field is the design
of powerful machines, such as functional entities that can
switch reversibly between distinct
geometries\cite{switching1,switching2,switching3,switching4}.
Such machines would not only be of great use on
their own but also could yield novel emergent behavior on
networking them together\cite{Whitesides}.
The existence of life in its myriad forms provides a proof-of-concept
of what one might aspire to accomplish with nanotechnology.

Proteins are complex water-soluble chain molecules
made up of tens or hundreds of twenty types of naturally
occurring amino acids and exhibit conformational switching\cite{Hunter}
triggered by influences such as ligand binding. At the
nanoscale, thermal fluctuations yield forces with magnitude
comparable to those involved in chemical reactions catalyzed by
the proteins\cite{SchrodingerBook}.
How might one design a machine whose functionality is
structure-based? In its simplest form, we seek an object which takes
on just a few distinct geometries in a reproducible manner and is
able to switch reversibly between them due to thermal fluctuations.
Such a situation would allow external stabilizing influences to
favor a given conformation over the others and allow for the
development of powerful machines at the nanoscale.

A collection of hard spheres constitutes the simplest model of
matter and exhibits both a crystalline phase and a fluid phase on
varying the density of spheres\cite{Chaikin}. A linear chain of hard
spheres is the simplest connected object with the fewest constraints
and thus the greatest flexibility\cite{Thorpe}. Such a chain,
at high temperatures or when there are no interactions promoting
compaction, would occupy a random coil phase in which all self-avoiding
conformations are equally likely. This situation is not conducive
for machine design. In the presence of inter-sphere interactions
promoting compaction and when there are no frustrating influences
from the sphere tethering, one would expect a generic compact phase
in which, at least locally, the preferred conformation is that of a
face-centered cubic (fcc) lattice. It was conjectured by Kepler, and
proven more recently by Hales, that the fcc structure provides for
optimal packing of unconstrained spheres\cite{Kepler}. There is
again a high degeneracy of compact conformations with no assurance that
any given conformation will be reached from a random coil conformation
rapidly and reproducibly. Again, this situation is not conducive for
the design of machines.

Of course, what one requires is a phase of matter with much fewer
ground state conformations than either the random coil or generic
compact phase. Additionally, one needs a phase which is in the
vicinity of a phase transition (or crossover for finite size
systems) to a distinct phase, because that provides for exquisite
sensitivity to the right types of external influences. Our thinking
is guided by the liquid crystal phase\cite{Chaikin} which is a
distinct state of matter that is poised in the vicinity of the
liquid phase. This phase is known to be one of the most sensitive
phases of matter. The liquid crystal phase opens up because of the
anisotropy of the constituent molecules -- there is no longer a need
for simultaneous ordering in all three directions. Rather, one can
have translational order in fewer than three dimensions and
orientational order. Here we address the issue of how one might open
up a distinct phase with relatively few ground state conformations
in the vicinity of the random coil phase. Such a phase would then be
the analog of a liquid crystal phase but this time for chain
molecules.

\begin{figure}
\centerline{\includegraphics[width=2.4in]{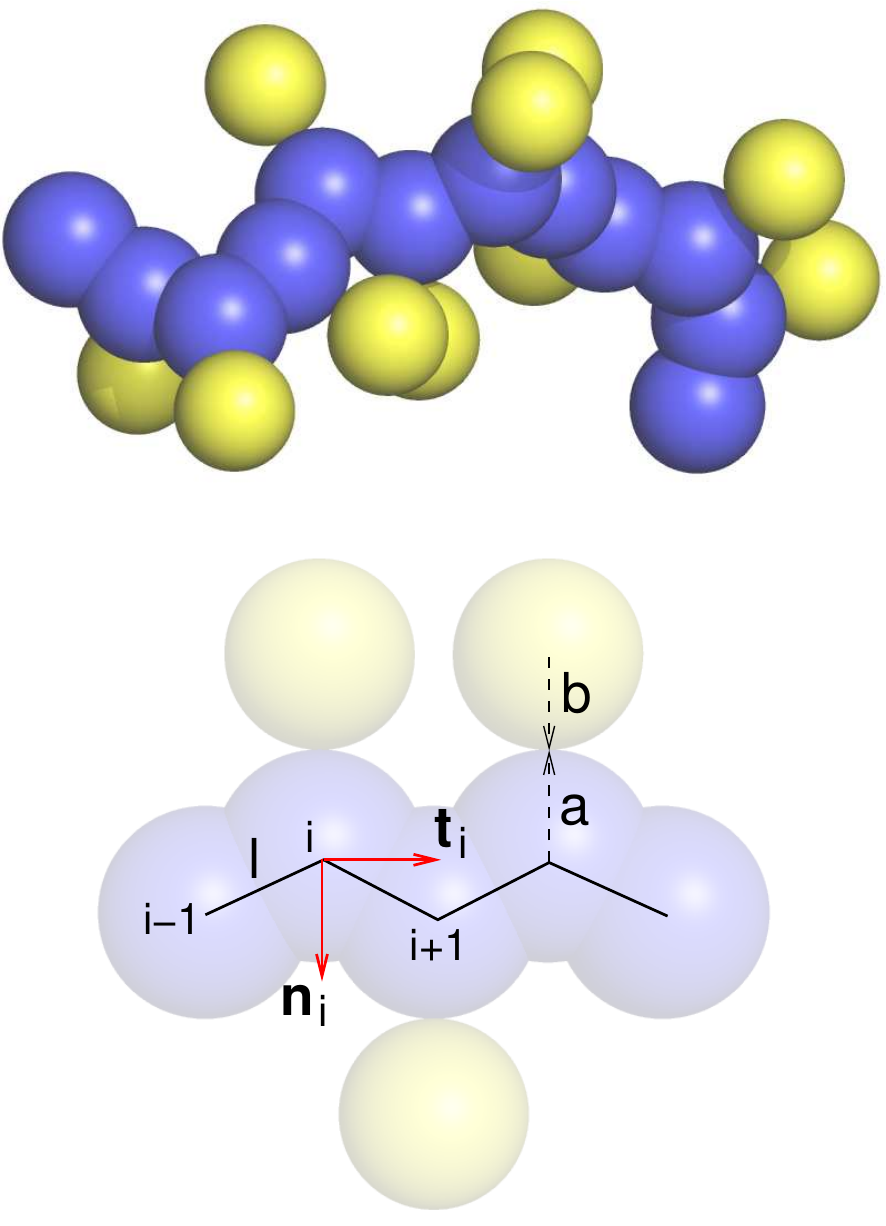}}
\caption{Sketch of the chain molecule. The backbone is modeled as a chain of
(blue) spheres of radius $a$ with a separation along the chain
$l=3.8\AA$. The nearest neighbor spheres along the backbone are
allowed to overlap with each other thereby overtly introducing
uniaxial anisotropy. ${\bf t}_i$ and ${\bf n}_i$ are the tangent and
normal vectors assigned to each sphere, $i$, except to
those at the ends of the chain. Side-spheres (shown in pink) of
radius $b$ are attached to the backbone spheres in the negative
normal direction. The side-spheres are not allowed to overlap with
either the backbone spheres or with each other.
}
\end{figure}

Armed with insights from the liquid crystal phase and the behavior of
proteins, we identify two mechanisms for thinning the number of
candidate ground state structures and demonstrate that these are
sufficient for the creation of the sought after behavior. One can
define a Frenet reference frame (a local Cartesian coordinate system)
at each location of the chain molecule comprised of the tangent, the
normal, and the binormal as the orthogonal coordinate axes.  These
coordinate systems play a crucial role in at least two ways, as can be
readily seen in the protein context: first, the chemistry of hydrogen
bonds and other chemical features yield constraints on the relative
orientations between the coordinate systems associated with pairs of
amino acids in contact; second, the side-chains of amino acids are
located in a specified direction with respect to these coordinate
systems - for example, the amino acid side chains are typically
pointed approximately in the outward normal direction. Indeed, both
the $\alpha$-helix and the $\beta$-sheet allow for the placement of
side-chains in a manner that avoids steric clashes.

\begin{figure}
\centerline{\includegraphics[width=3.2in]{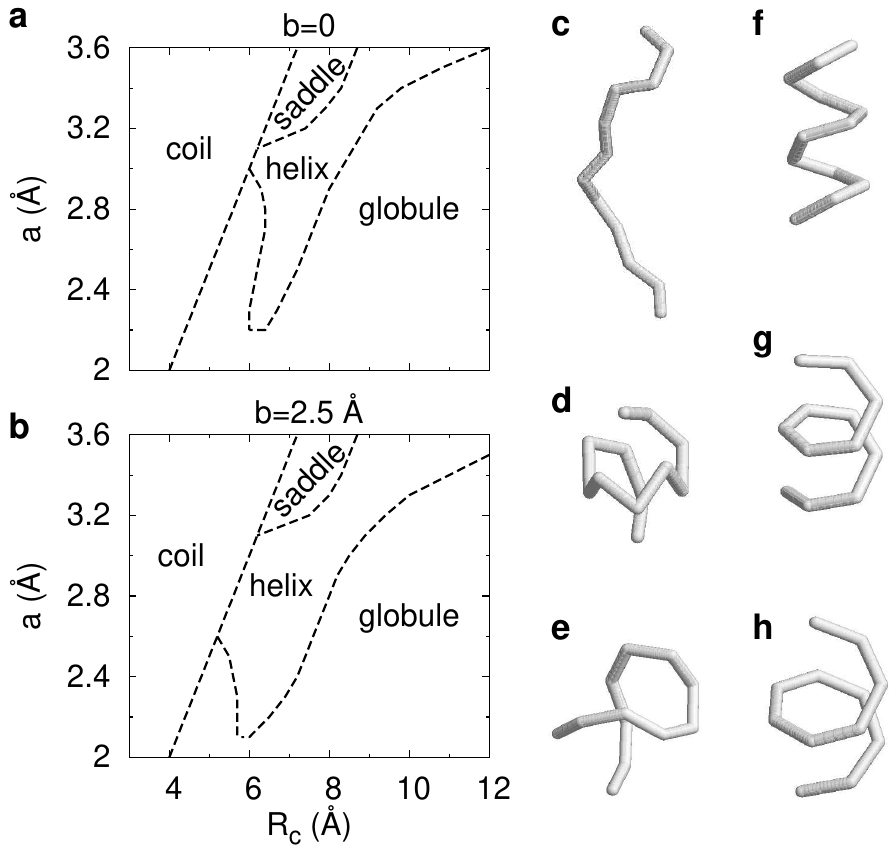}}
\caption{Phase diagram of chain molecules made up of
spheres. The panels ({\bf a}, {\bf b}) show the phase
diagrams at $T =0$ in the $a - R_c$ plane for chains of
length $N$=12 without side-spheres ($b=0$) and with
side-spheres ($b=2.5 \AA$) respectively. The intermediate
compact phase arises on the edge of compaction of the
chain molecule when $R_c$ becomes sufficiently large to
allow the attractive interaction to be effective. The
vicinity of this phase to other phases (it is
sandwiched between two distinct phases) confers
sensitivity to structures in the intermediate compact
phase. The other panels show typical ground state
conformations: a random coil conformation obtained with
$R_c < 2a$ ({\bf c}), a globule conformation obtained
with $a=2.5\AA$, $R_c=9\AA$ ({\bf d}), a saddle
conformation obtained with $a=3.2\AA$, $R_c=7\AA$ ({\bf
e}), several helix conformations with different pitch
to radius ratios obtained for $a=2.6\AA$, $R_c=6.4\AA$
({\bf f}), $a=3.2\AA$, $R_c=8\AA$ ({\bf g}), and
$a=3.4\AA$, $R_c=9\AA$ ({\bf h}). The helical and
saddle ground state conformations in the model with no
side-spheres are retained even when side-spheres are
present ($b=2.5\AA$). However, the side-spheres
eliminate several random globule and random coil
conformations thereby stabilizing the intermediate
compact phase.
}
\end{figure}

The first mechanism follows from the observation that a model of a
chain comprised of spheres is unable to capture the inherent
anisotropy induced by the presence of the chain constraint -- at each
location of the chain, there is a tangent direction defined by the two
adjacent objects. Thus the simplest model capturing the correct
symmetry is one in which the constituent monomers are no longer
\textbf{isotropic}. We therefore allow for the overlap of van der Waals
spheres of adjacent monomers along the chain. Such an overlap overtly
breaks the isotropy of the spheres and confers uniaxial anisotropy to
the chain.  In both the emergent building blocks of protein
structures, $\alpha$-helices and $\beta$-sheets, nearby segments are
placed right up against each other and aligned parallel to each other,
reflecting the anisotropy. In a protein, the distance between
neighboring C$^{\alpha}$ atoms is $3.8\AA$ requiring that the radius
of the backbone of an amino acid be larger than $1.9\AA$.  This
constraint is easily met by considering the van der Waals radii of the
constituent atoms. For example, one would estimate\cite{VdWAA} that
the smallest amino acid, glycine, has a radius larger than $2.4\AA$.
Second, we study the effect of attaching side-chains to the backbone
monomer, e.g., in the negative normal direction. The self-avoidance of
these side spheres with each other and with the backbone spheres
results in an induced coupling between pairs of Frenet frames.

\begin{figure}
\centerline{\includegraphics[width=3in]{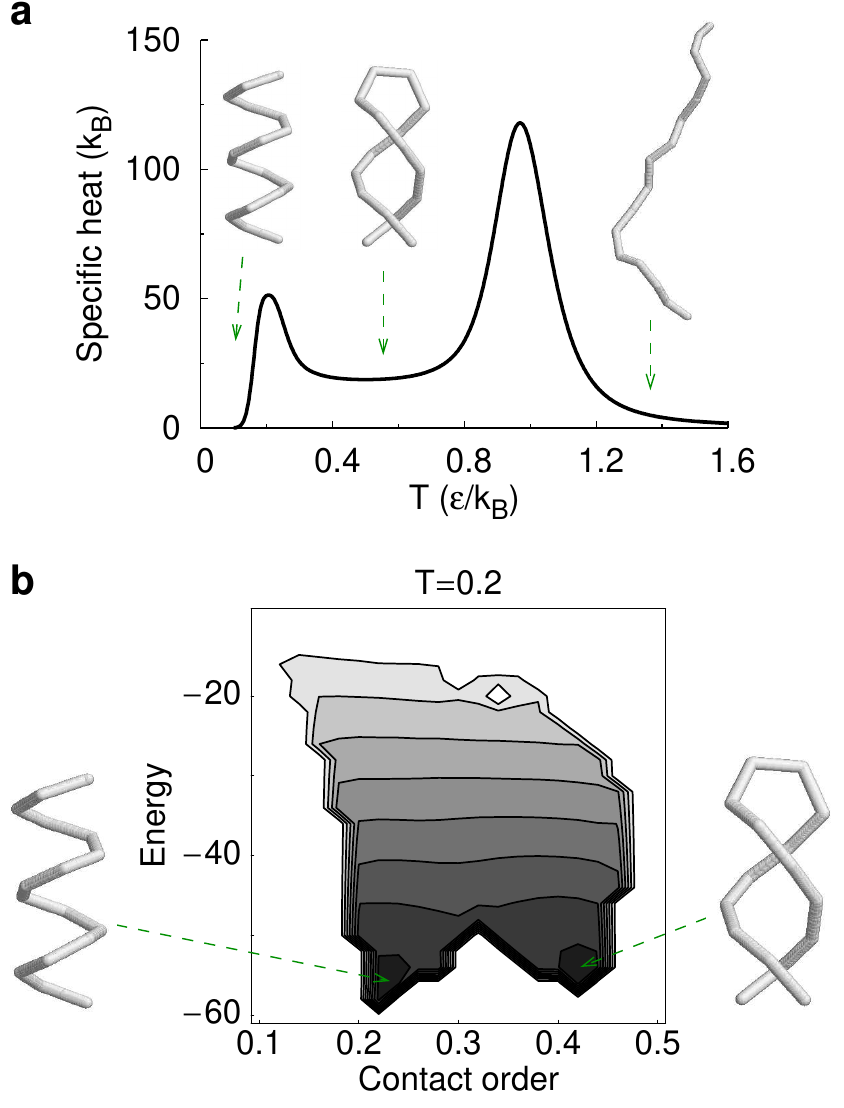}}
\caption{Multiple specific heat peaks corresponding
to three distinct phases. We study a chain of $N=16$
with $a=3\AA$, $b=2.5\AA$, and $R_c=7.5\AA$.
({\bf a}) The temperature dependence of the specific
heat indicates two peaks -- one at $T\approx 0.2
\varepsilon/k_B$ and another at $T\approx 0.97
\varepsilon/k_B$. The lower temperature peak indicates
a crossover into a single helix conformation from a
dual helix, whereas the peak at the higher
temperature indicates a crossover between the dual
helix and the random coil phases. ({\bf b}) Contour plot
of the effective free energy as a function of energy
and contact order, $CO$, at the lower transition
temperature. $CO$, is defined as the sum over sequence
separation of all contacts divided by the product of
the number of contacts and $N$. The effective
free energy at a given temperature $T$ is determined as
${\cal F}(E,CO) = -k_B T \ln P(E,CO)$, where $P(E,CO)$
is a weighted two-dimensional histogram for that
temperature obtained using the multiple histogram
method\cite{Ferrenberg}.
The unweighted histograms at
multiple temperatures are collected through parallel
tempering\cite{tempering}
Monte Carlo simulations. The free energy
difference between consecutive contour levels is
$4\varepsilon$.  The contour plot shows two minima
corresponding to the single helix and the dual helix
as indicated.
}
\end{figure}

We find that each of these features results in the creation
of a distinct intermediate phase for short chains sandwiched between
the generic compact phase and the random coil phase. Furthermore, when
the two features are combined, the phase is stabilized and occupies a
larger region in parameter space. For our simplified model of a chain
of backbone spheres (Fig. 1) with a sphere separation along the
sequence of $3.8\AA$, we have three length parameters: the radius,
$a$, of the backbone spheres, the radius, $b$, of the side-spheres,
and the cut-off scale, $R_c$, for the pairwise attractive interaction
potential, $\varepsilon/k_B$, between the backbone spheres (See
Materials and Methods).
In the absence of side-spheres and for
non-overlapping backbone spheres, $a \leq 1.9\AA$, there is a
crossover from a random coil phase (in which the conformations are
mostly extended) to a globule phase in which one has compact
conformations with no distinct motifs such as helices.  The situation
becomes qualitatively different on incorporating one or both of the
two key features.

Fig. 2a shows the $T=0$ phase diagrams in the $a$--$R_c$
plane for a $N=12$ chain. Distinct phases, corresponding to
helical and saddle-like conformations, emerge between the coil
and the globular phases analogous to the opening up of a liquid
crystal phase between the liquid and crystalline phases for
anisotropic molecules\cite{Chaikin}.  The saddle can be thought
of as a piece of a dual helix structure and does develop into a
dual helix for longer chains. (The dual helix is distinct from the
double helix because the former is a conformation of a single chain
whereas the latter is comprised of two chains.)
These phases have a lower entropy
than both the globule and random coil (swollen) phases but are
stabilized by the attractive interaction potential. We also find that
the presence of side-spheres even without overlap of the backbone spheres
induces helical conformations (Figure S1 in SI).
This effect is accentuated when both features are present simultaneously
(Figure S2 in SI). Interestingly, conformations
composed of nearly parallel strands similar to those in $\beta$-sheets
can be obtained by adding a bending energy into the model
(Figure S3 in SI).

We now consider an even shorter system ($N=16$) with an
enhanced overlap of the backbone spheres: $a=3.0\AA$. The
radius of the side-sphere is chosen to be $b=2.5\AA$, and the
range of attraction is now increased to $R_c=7.5\AA$. These
parameter values yield a helix ground state but is close to a
crossover to the dual helix state. Fig. 3a shows that the
specific heat has two peaks corresponding to two crossover
temperatures, $T_1\approx 0.2\varepsilon/k_B$ and
$T_2\approx0.97\varepsilon/k_B$. At the lowest temperatures,
the helix is the dominant conformation. At intermediate
temperatures, one obtains the dual helix structure, whereas,
at higher temperatures, one obtains a random coil. Fig. 3b
shows the contour plots of the free energy on the
energy--contact order plane at $T_1$ and exhibits pronounced
minima corresponding to the single helix and dual helix.  The
helix is a bit lower in energy while the dual helix is
entropically more favorable. The free energy barrier for the
switching at $T_1$ is about $4\varepsilon$ which is equivalent
to the energy required to break 4 contacts. This barrier is
significantly higher than $k_B T_1$ and the two helical
conformations are each quite stable at this temperature. Quite
remarkably, our Monte Carlo simulations with standard pivot and
crankshaft move sets show a dynamical switching between the two
conformations in a single trajectory at a somewhat higher
temperature of $T=0.5\varepsilon/k_B$. The system is bistable
with a rapid switching between distinct conformations (for
examples of molecular switches,
see Refs.\cite{switching1,switching2,switching3,switching4}) with
little weight for intermediate conformations (Fig. 4).
Note that the switching can also be easily effected by means of an external
influence which is sensitive to the chain end-to-end distance.

It is important to note that the ease with which one obtains this
system without any fine-tuning of details is made possible by the
existence of the intermediate phase. The conformations in this
phase are in the vicinity of the random coil phase while retaining
order due to the proximity of the compact phase.
This results in special sensitivity to small perturbations induced,
e.g., by thermal fluctuations which, in the example presented here,
are responsible for the switching between two distinct geometrical
shapes. Unlike the random coil conformations which can switch from
one to another easily due to thermal fluctuations, the structures in
the intermediate phase exhibit some stability. At the same time, the
structures are not so densely compact that they are subject to
sluggish dynamics and kinetic inaccessibility characteristic of the
glassy phase. These distinct advantages of the intermediate phase have a
wider applicability than for chain molecules as evidenced by the
sensitive liquid crystal phase sandwiched between the crystalline
and liquid phases. In liquid crystals, the anisotropy arises from
the asymmetric shape of the constituent molecules, whereas here the
anisotropy is a natural consequence of the chain topology of the
molecule. Interestingly, earlier computational studies\cite{Clementi,Magee}
had found that overlapping adjacent monomers yield helical
conformations. Also protein-like folds are adopted by a host of
non-biological polymers\cite{Frauenrath,Greer,Greer1,Yang,Vincent,Sanda,Gong}
(see Appendix).

Understanding the properties of matter is greatly simplified on
using the concept of its phases\cite{Chaikin}. For example, a liquid
possesses certain gross properties, such as adopting the shape of
the container and its ability to flow, irrespective of its
constituent molecules and their underlying chemistry. Globular
proteins share a great deal of common characteristics - they are all
linear chains of the same 20 amino acids, they fold rapidly and
reproducibly into their native state structures, these structures
are made up of emergent building blocks in the form of helices and
almost planar sheets, the total number of distinct folds that
proteins adopt is limited in number\cite{Chothia,Chothia2} much as
the number of space groups associated with Bravais
lattices\cite{Chaikin} is $230$, proteins are flexible and versatile
in their folded state, and proteins have a tendency to aggregate and
form amyloid which in turn is implicated in debilitating human
diseases. These common attributes of proteins suggest that protein
structures occupy an intermediate phase of matter\cite{BANAVARMARITAN-ANN.REV.,HoangPNAS}
which confers on
them their many amazing characteristics.

\begin{figure}
\centerline{\includegraphics[width=3in]{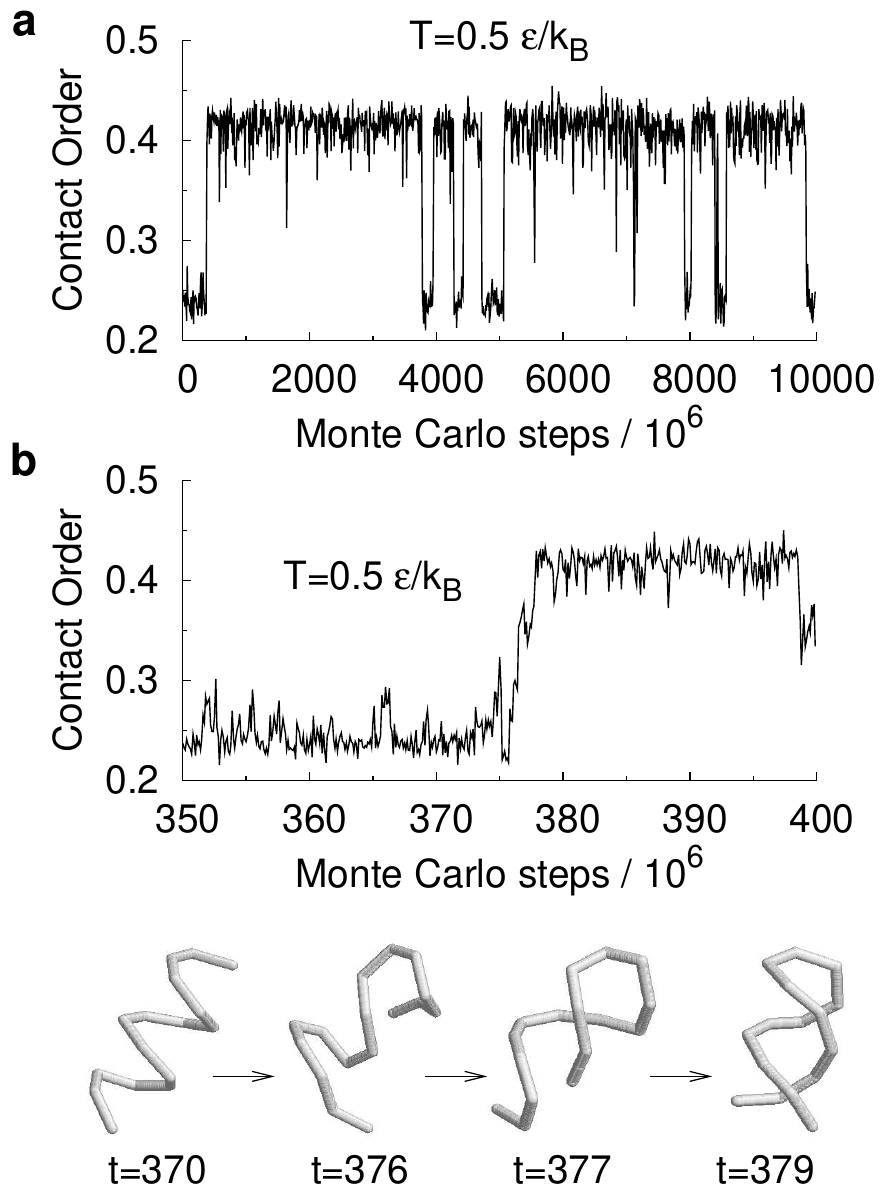}}
\caption{Dynamical switching between the single helix and the dual helix.
({\bf a}) A long Monte Carlo trajectory, with standard pivot and
crankshaft move sets\cite{Sokal},
at $T=0.5\varepsilon/k_B$, shows frequent
switches between the single helix (lower contact order) and the
dual helix (higher contact order). The chain studied here is the
same as in Fig. 3. Both the single and the dual helix have
similar energies. ({\bf b}) Similar to panel {\bf a} but for a much
smaller window of time steps and with higher resolution showing a
single switch from the single helix to the dual helix. The bottom
figures are the snapshots during the switch at times $t$ (in units
of $10^6$ steps) as indicated.
}
\end{figure}

Whitesides wrote\cite{Whitesides}
that one ought to ``take existing
nanomachines--those present in the cell--and learn from them. We
will undoubtedly be able to extract from these systems concepts and
principles that will enable us to make variants of them that will
serve our purposes, and others that will have entirely new
functions". The phase of matter so successfully used by nature as
the basis of life is ready to be exploited in the laboratory.

\subsection*{Methods}
We consider chains of $N$ hard spheres each of radius $a$.
The beads spacing along the chain is fixed to be $l=3.8\AA$.
Non-consecutive spheres are not allowed to overlap, whereas the
neighboring spheres along the chain can overlap when $a$ is larger
than $1.9\AA$. Non-consecutive spheres interact via a pairwise
square-well potential equal to $-\varepsilon$ within a contact
range, $R_c$. A Frenet frame of reference is assigned to each
sphere, $i$, except to those at the ends of the chain. The tangent
vector, $\hat{\bf t}_i$, in this frame is a unit vector tangential
to the circle passing through the centers of beads $i-1$, $i$ and
$i+1$. The Frenet normal vector, $\hat{\bf n}_i$, is an unit vector
pointing to the center of this circle. The third vector of the frame
is the binormal vector, denoted by $\hat{\bf b}_i$, and is defined
to be the cross product of the tangent and the normal vectors. A
side-sphere of radius $b$ is attached to each backbone sphere in
direction opposite to the normal with the distance between the
centers of the backbone sphere and its side-sphere partner equal to
$a+b$. The backbone spheres at the two ends of the chain do not have
any side-spheres attached to them. The role of the side-spheres is
entirely steric - they are not allowed to overlap with any of the
other spheres in the system. The energy of the chain in 
given conformation can be written as
\begin{equation}
E = -\varepsilon \sum_{i=1}^{N-2} \sum_{j=i+2}^N
\Theta(R_c - |{\vec r}_i - {\vec r}_j|) \;,
\end{equation}
where ${\vec r}_i$ are center positions of the backbone spheres, 
$N$ is the number of such spheres, and the step function $\Theta(x)$ 
is equal to 1 if $x>0$ and 0 otherwise.

We employ a parallel tempering\cite{tempering} Monte Carlo (MC)
scheme for obtaining the ground state as well as other
equilibrium characteristics of the system. The simulation
entails monitoring 20-30 replicas, each evolving at its own
selected temperature, $T_i$. For each replica, the simulation
is carried out with the standard pivot and crankshaft move
sets\cite{Sokal} and the Metropolis algorithm for move acceptance.
In a pivot move, one randomly chooses a sphere $i$ in the chain
and rotates the shorter part of the chain (either from 1
through $i-1$ or from $i+1$ through $N$, where $N$ is the
number of spheres in the chain) by a small angle and about a
randomly chosen axis that goes through the $i$-th sphere. In a
crankshaft move, two spheres $i$ and $j$ are chosen randomly
such that $|i-j|<6$ and the spheres between $i$ and $j$ are
rotated by a small angle and about the axis that goes through
$i$ and $j$. In both move sets, the rotation angle is randomly
drawn from a Gaussian distribution of zero mean and a
dispersion of $4^o$.
An attempt to exchange replicas $i$ and $j$ is made every 100
MC steps. The exchange is accepted with a probability equal to
$p_{ij} = \max\left\{1,\exp[k_B^{-1}(T_i^{-1} - T_j^{-1})(E_i -
E_j)] \right\}$, where $k_B$ is the Boltzmann constant, and
$E_i$ and $E_j$ are the energies of the replicas at the time of
the exchange. The weighted multiple histogram
technique\cite{Ferrenberg} is used to compute the specific heat
and the effective free energy.

\subsection*{Appendix: Protein-like folds are adopted by a host of
non-biological polymers}
Poly(diacetylene)s can form
multiple-helical superstructures \cite{Frauenrath}
such as
double-helical ribbons. Poly(ethylene glycol) (PEG) has been shown
to assume a helical conformation in isobutric acid with a trace
amount of water without which PEG forms a coil
configuration\cite{Greer,Greer1}.
PEG also forms helices in isopentanoic and
n-propanoic acids but not in isobutanol or n-butanol\cite{Greer1}.
In isobutyric acid, PEG forms a mixture of helices and coils whereas
a similar polymer poly(ethy-lene imine) (PEI)  merely forms helices.
Phenylacetylene oligomers of specific chain lengths (up to 18 units)
and containing a tri(ethylene oxide) side-chain segment at each
repeat unit have been found to undergo sharp switching between
arrays of random coils and arrays of helical conformations on
changing the solvent composition\cite{Yang}.
Another interesting
example is provided by poly(ethylene oxide) (PEO) dissolved in an
electrolyte consisting of a lithium salt\cite{Vincent},
LiCF$_3$SO$_3$.
In the crystalline phase of the system, the PEO
chains are helical and form parallel arrays. In the amorphous phase,
the arrays dissolve into separate helices. Finally, there are
several examples of synthesis of helical polymers that are discussed
by Sanda et al.\cite{Sanda}
and which include polychloral,
polyisocyanates, polyisocyanides, polisilanes, and polycetylenes.
Such helical structures can be made to have adjustable geometry and
then used to generate nanocavities of tunable sizes\cite{Gong}.
Sanda et al.\cite{Sanda}
have recognized that the crucial factor
for the synthetic polymers to adopt the helical structures is the
steric repulsion between the side chains combined with attraction
that is usually provided by hydrogen bonding.

\begin{acknowledgments}
This work was supported by PRIN 2007, NAFOSTED, 
the grant N N202 0852 33 from the Ministry of Science and Higher
Education in Poland, and the Vietnam Education Foundation (VEF).
T.X.H. is a VEF Visiting Scholar.
\end{acknowledgments}

\section*{Supporting Information}

\begin{center}
\includegraphics[width=3.4in]{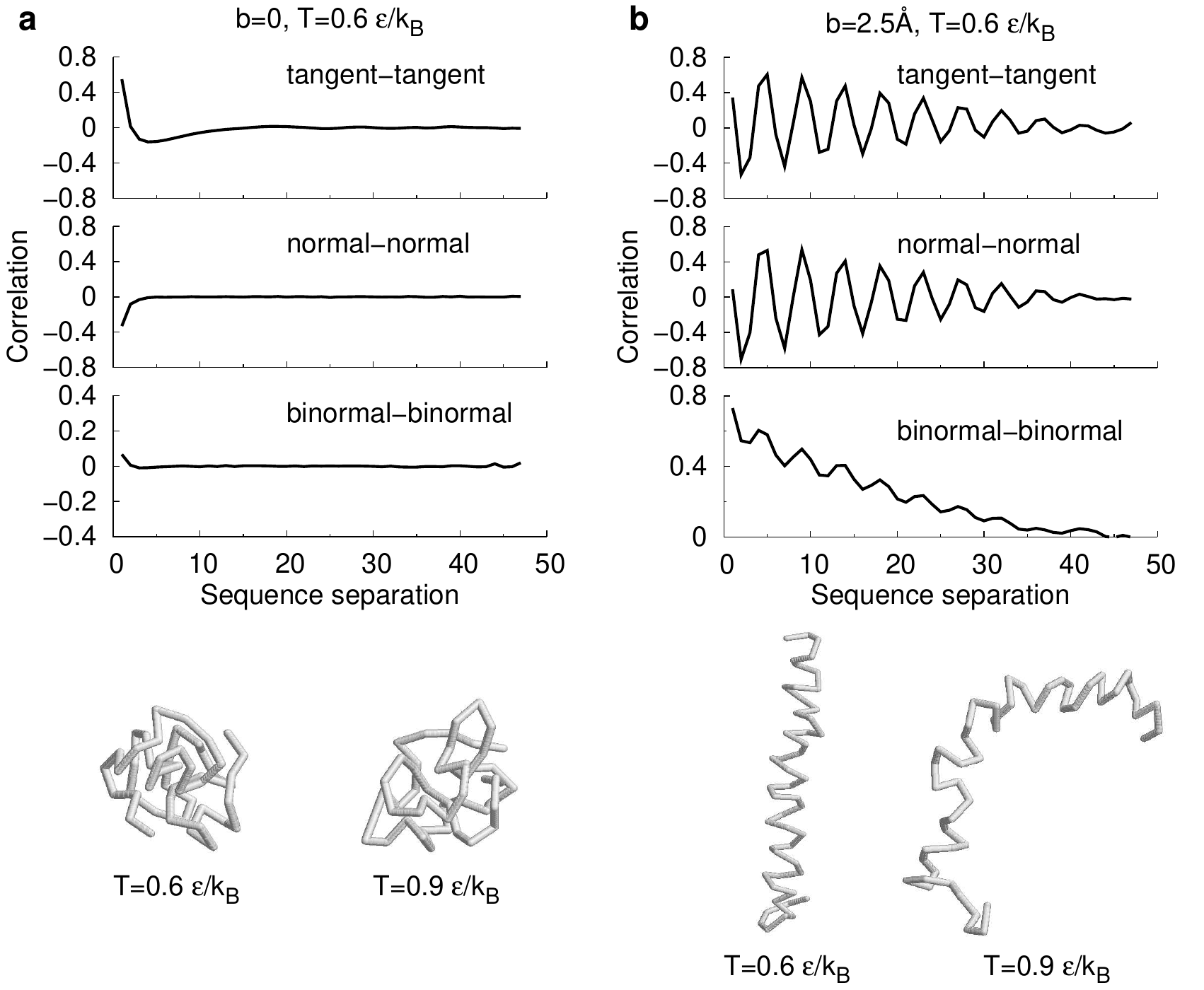}
\end{center}
Figure S1:
Correlation functions for a chain of non-overlapping spheres.
The tangent-tangent, normal-normal, and the
binormal-binormal correlations as a function of sequence separation
are shown for a chain of $N=50$ non-overlapping spheres
($a=1.9\AA$) with $R_c=6.5\AA$. The chain is studied at
$T=0.6 \varepsilon/k_B$ which is below the collapse transition.
Panel {\bf a} corresponds to the situation without the side-spheres
($b=0$)  and panel {\bf b} with side-spheres ($b=2.5\AA$). The
tangent-tangent correlation for a sequence separation $s$ is defined
as $\left<\hat{\bf t}_i\cdot\hat{\bf t}_{i+s}\right>$ where the
average $\left<\cdot\right>$ is taken over all $i$ running from 2 to
$(N-s-1)$ and over all sample conformations. The normal-normal and
binormal-binormal correlations are defined similarly.  The
correlations are averaged over a run of $10^9$ MC steps, in which
conformations are considered every $10^5$ steps. The conformations
displayed at the bottom are typical snapshots obtained in the
simulation at two different temperatures as indicated. The two
conformations on the left are for $b=0$ and the remaining two are
for $b=2.5\AA$. Note that the presence of the
side-spheres induce helical conformations and yield
qualitatively distinct behaviors of the correlation functions
of the tangent, normal, and binormal vectors along the chain.

\clearpage

\begin{center}
\includegraphics[width=3.4in]{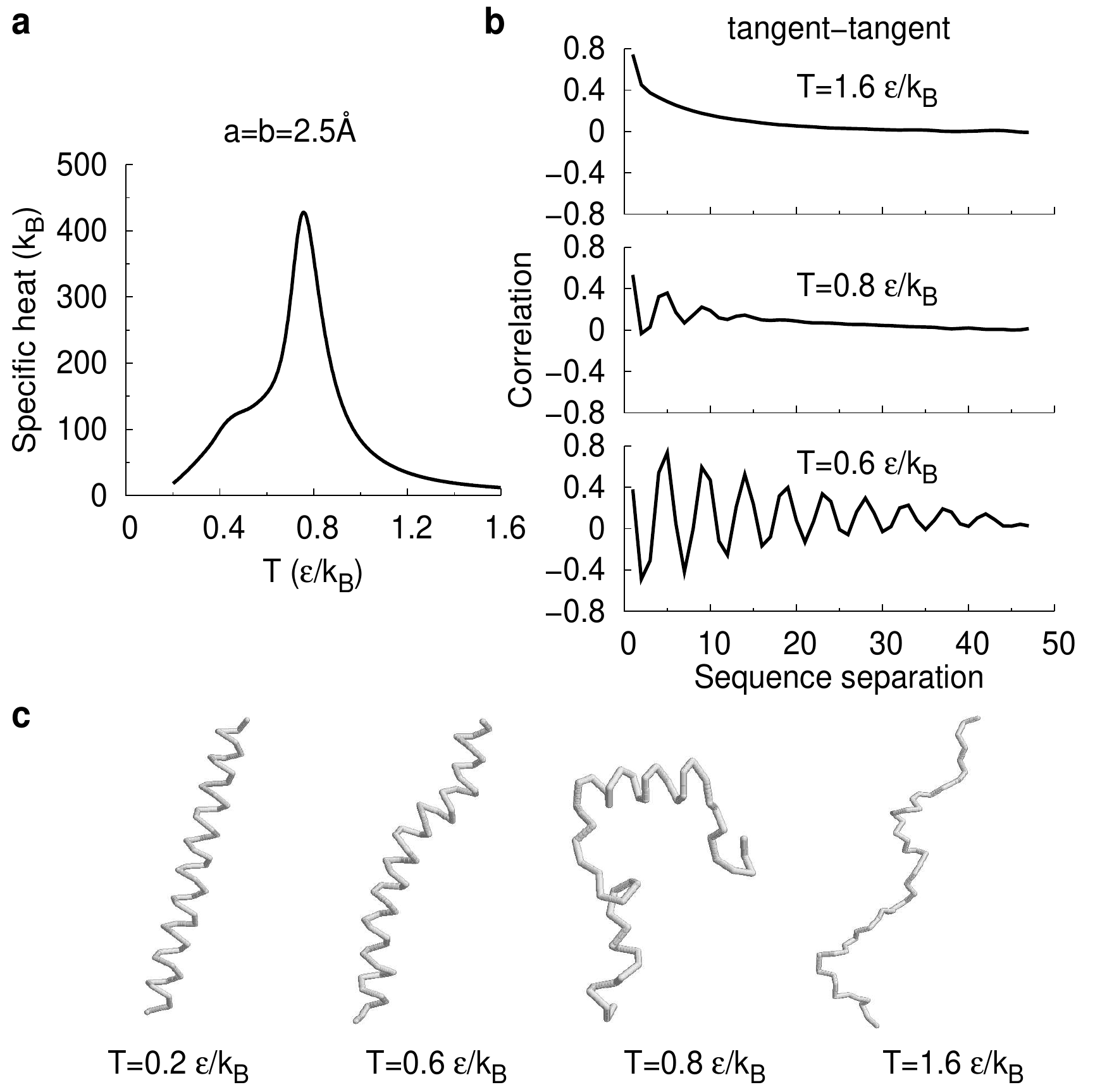}
\end{center}

Figure S2:
The crossover between the single helix and the random coil phases
in a chain of $N=50$ overlapping spheres. The parameters are
$a=2.5\AA$, $b=2.5\AA$, and $R_c=6.5\AA$. The ground state for
this system is a single helix.  ({\bf a}) Dependence of the
specific heat on temperature showing a peak at $T_{max} \approx
0.72 \varepsilon/k_B$.  ({\bf b}) Dependence of the
tangent-tangent correlation on the sequence separation for
three different temperatures: well-above the transition
($T=1.6\varepsilon/k_B$), just above the transition
($T=0.8\varepsilon/k_B$) and just below the transition
($T=0.6\varepsilon/k_B$).  ({\bf c}) Snapshots of typical
conformations at the three temperatures. The specific heat is
obtained from fluctuations in the energy using the equation $C =
(\left<E^2\right>-\left<E\right>^2)/(k_B T^2)$, where $E$ is the
energy and $\left< \cdot \right>$ indicates an average.
Note that the signature of the helical ground
state is present in the correlation functions even at a temperature
distinctly higher than $T_{max}$ showing that information about the
ordered low temperature state is encoded even in the high
temperature phase.

\begin{center}
\includegraphics[width=2.6in]{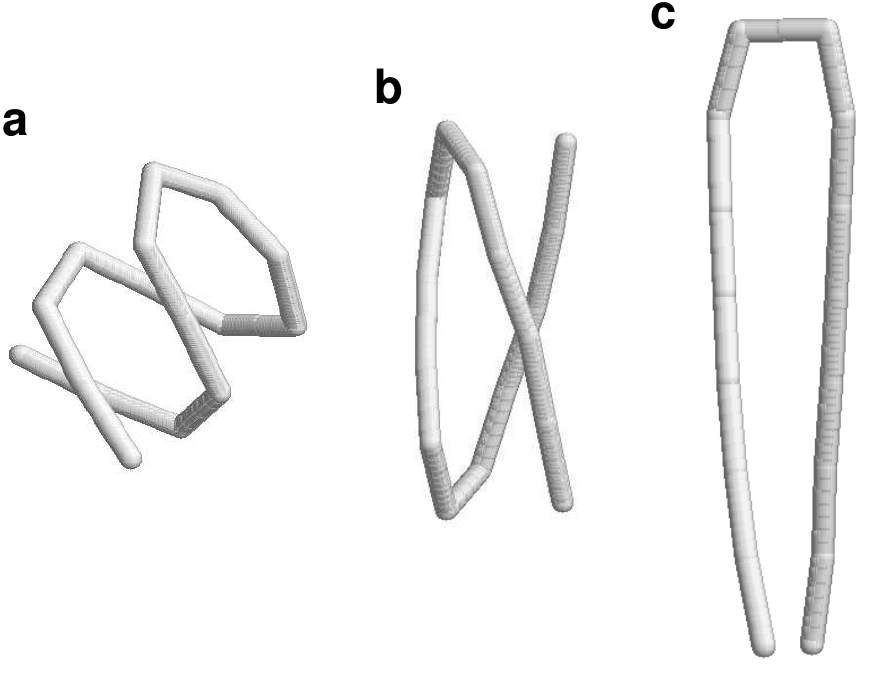}
\end{center}

Figure S3:
Ground state conformations of a chain of $N=16$ overlapping spheres with
side-spheres and an additional bending energy along the chain. The
parameters are $a=2.5\AA$, $b=2.5\AA$ and $R_c=6.5\AA$. The bending energy
is given by $\sum_{i=2}^{N-1} e_R(1-\cos(\theta_i))$, where $e_R$ is a
coefficient and $\theta_i$ is the angle between two consecutive
connectivity vectors ${\bf r}_{i-1,i}\equiv{\bf r}_{i}-{\bf r}_{i-1}$ and
${\bf r}_{i,i+1}\equiv{\bf r}_{i+1}-{\bf r}_{i}$ associated with the
sphere $i$. The conformations shown are for $e_R=2\varepsilon$ (a),
$4\varepsilon$ (b) and $8\varepsilon$ (c), respectively. Note that for the
largest value of $e_R$, one obtains a planar hairpin-like conformation.

\end{document}